\pdfoutput=1

\documentclass[twocolumn]{aastex631}

\newcommand{\citeg}[1]{\citep[e.g.,][]{#1}} 
\newcommand{\citesa}[1]{\citep[see also][]{#1}} 

\shorttitle{}
\shortauthors{Utsumi et al.}
\graphicspath{{./}{figures/}}
\usepackage{tensor}
\usepackage{color}
\usepackage[T1]{fontenc} 

\usepackage{CJKutf8}

\begin{document}
\begin{CJK}{UTF8}{ipxm}

\title{
Component of energy flow from supercritical accretion disks around rotating stellar mass black holes.
}

\correspondingauthor{Aoto Utsumi}
\email{utsumi@ccs.tsukuba.ac.jp}

\author[0000-0003-3208-7094]{Aoto Utsumi}
\affiliation{University of Tsukuba, 1-1-1 Tennodai, Tsukuba, Ibaraki 305-8577, Japan}

\author[0000-0002-2309-3639]{Ken Ohsuga}
\affiliation{University of Tsukuba, 1-1-1 Tennodai, Tsukuba, Ibaraki 305-8577, Japan}

\author[0000-0003-0114-5378]{Hiroyuki R. Takahashi}
\affiliation{Komazawa University, 1-23-1 Komazawa, Setagaya, Tokyo 154-8525, Japan}

\author[0000-0003-3640-1749]{Yuta Asahina}
\affiliation{University of Tsukuba, 1-1-1 Tennodai, Tsukuba, Ibaraki 305-8577, Japan}



\begin{abstract}
By performing two-dimensional axisymmetric general relativistic radiation magnetohydrodynamics simulations with spin parameter $a^*$ varying from -0.9 to 0.9, we investigate the dependence on the black hole spin of the energy flow from supercritical accretion disk around stellar mass black hole.
It is found that optically and geometrically thick disks form near the equatorial plane, and a part of the disk matter is launched from the disk surface in all models.
The gas ejection is mainly driven by the radiative force, 
but magnetic force cannot be neglected, when $|a^*|$ is large.
The energy outflow efficiency
(total luminosity normalized by $\dot{M}_{\rm in} c^2 $; $\dot{M}_{\rm in}$ and $c$ are the mass accretion rate at the event horizon and the light speed)
is larger for rotating black holes than for non-rotating black holes.
This is $0.7\%$ for $a^*=-0.7$, 
$0.3\%$ for $a^*=0$, and $5\%$ for $a^*=0.7$
for $\dot{M}_{\rm in} \sim 100L_{\rm Edd}/c^2$
 ($L_{\rm Edd}$ is Eddington luminosity).
Also, although the energy is mainly released by radiation
when $a^* \sim 0$, the Poynting power
increases with $|a^*|$ and exceeds the radiative luminosity
for models with $a^* \geq 0.5$ and $a^* \leq -0.7$.
The more the black hole rotates, the larger the power ratio of the kinetic luminosity to the isotropic luminosity tends to be.
This implies that objects with large (small) power ratio may have rapidly (slowly) rotating black holes.
Among ultraluminous X-ray sources, IC342 X-1, is a candidate with a rapidly rotating black hole.
\end{abstract}

\keywords{accretion, accretion disks --- magnetohydrodynamics (MHD) --- radiation: dynamics --- stars: black holes}


\section{Introduction} \label{sec:intro}
Luminous compact objects such as X-ray binaries, microquasars, active galactic nuclei, and gamma-ray bursts are considered to be powered by the gas accretion on black holes. 
Black hole accretion is known to exhibit several different modes. 
If the mass accretion rate comparable to or less that the Eddington limit ($L_{\rm Edd}/c^2$ where $L_{\rm Edd}$ and $c$ being the Eddington luminosity and speed of light), geometrically thin and optically thick disk (so-called standard accretion disk) is formed \citep{shakuraBlackHolesBinary1973}. 
When the mass accretion rate is much smaller than the Eddington limit, radiatively inefficient accretion flows (RIAF), which consists of the hot rarefied plasma appear \citep{narayanAdvectiondominatedAccretionSelfsimilar1994,narayanAdvectiondominatedAccretionSelfSimilarity1995,narayanAdvectiondominatedAccretionUnderfed1995,abramowiczThermalEquilibriaAccretion1995}.
Supercritical accretion flows, on the other hand, are geometrically and optically thick disks, whose luminosity exceeds the Eddington luminosity and form when the accretion rate exceeds the Eddington limit.

The supercritical disks are thought to be the energy source for extremely bright compact objects. Ultraluminous X-ray sources (ULXs) are off-nuclear sources with X-ray luminosities of $\gtrsim 10^{39} \rm erg\,s^{-1}$ and may be undergoing supercritical accretion into stellar mass black holes. 
Other possibilities have been suggested as subcritical accretion onto intermediate-mass black holes and supercritical accretion onto neutron stars. For the ULX, for which X-ray pulses have been detected, a neutron star is believed to exist in the center, but the central objects of the other ULXs are unknown. 
In this study, we investigate the case of supercritical accretion onto the stellar mass black hole.

Radiation hydrodynamics (RHD) is necessary in the study of supercritical accretion because the interaction between the radiation field and the material is important. 
\citet{ohsugaSupercriticalAccretionFlows2005b} was the first to succeed in reproducing the steady-state of supercritical accretion flow using RHD simulations \citesa{eggumJetProductionSuperEddington1985,eggumRadiationHydrodynamicCalculation1988}.
Subsequently, the radiation magnetohydrodynamics (RMHD) simulations were developed 
in order to account for the effects of magnetic fields, which are the origin of disk viscosity \citeg{ohsugaGlobalRadiationMagnetohydrodynamicSimulations2009,ohsugaGlobalStructureThree2011,jiangGLOBALTHREEDIMENSIONALRADIATION2014}.
It has been reported by these simulations that a part of the disk matter is blown away by the strong radiation pressure force. The bubble structure detected in some ULXs \citep{pakullOpticalCounterpartsUltraluminous2002,pakullBubbleNebulaeUltraluminous2003,griseUltraluminousXraySource2006,abolmasovSpectroscopyOpticalCounterparts2007,csehBlackHolePowered2012} may have been formed by these radiatively driven outflows \citep{hashizumeRadiationHydrodynamicsSimulations2015}.

However, a non-rotating black hole is supposed in these simulations
and supercritical accretion on a rotating black hole is not yet well understood 
even though the black hole may be rotating.
ULXs would be kind of X-ray binaries \citep{komossaROSATViewNGC1998a}, but there is a possibility of an isolated black hole that plunges into an interstellar matter \citep{miiULTRALUMINOUSXRAYSOURCES2005,kobayashiNewPossibleAccretion2019}.
In the latter case, the rotation axis of the black hole can be misaligned with that of the disk. 
It is also possible that the rotation direction of the black hole and the rotation direction of the disk are opposite.
It is known that the SS433 has a precessed jet \citep{cramptonProbableBinaryNature1980,cherepashchukSS433Eclipsing1981}
and one possible interpretation of this is a misaligned system.

In order to investigate the effect of black hole spin, it is necessary to perform RMHD simulations that incorporate general relativistic effects, i.e., general relativistic (GR) RMHD simulations.
GR-RMHD simulations of supercritical accretion flows around the non-rotating black hole have been performed by \citet{sadowskiNumericalSimulationsSupercritical2014b} \citesa{sadowskiPowerfulRadiativeJets2015a}.
\citet{sadowskiNumericalSimulationsSupercritical2014b} has been carried out the simulation for the case of a rotating black hole, but only the spin parameter of 0.9 is employed \citesa{takahashiFORMATIONOVERHEATEDREGIONS2016}.

In the present study, we investigate the effect of the black hole spin on the supercritical accretion flows around the black holes by performing two-dimensional GR-RMHD simulations with spin parameter $a^*$ varying from $-0.9$ (retrograde) to $0.9$ (prograde). 
The disk luminosity, the power of the ejected gas, their dependence on the spin parameter
in the quasi-steady state 
are investigated and compared with the observations of some ULXs. 
The paper is organized as follows: we describe the methods of
calculations and initial conditions in Section \ref{sec:basic}.
In Section \ref{sec:res}, we show our simulation results.
Section \ref{sec:dis} is devoted to discussion.
This section contains comparisons with observations of ULXs.
We also discuss about the magnetically arrested disks (MADs)
\citeg{igumenshchevThreedimensionalMagnetohydrodynamicSimulations2003,narayanMagneticallyArrestedDisk2003,tchekhovskoyEfficientGenerationJets2011a,tchekhovskoyGeneralRelativisticModeling2012c,mckinneyGeneralRelativisticMagnetohydrodynamic2012b},
since our present study focuses on the standard and normal evolution (SANE) disks and does not deal with the MADs.
Finally, we summarize our main findings in Section \ref{sec:sum}.

\section{Basic Equations and Numerical Method} \label{sec:basic}
We numerically solve GR-RMHD equations using the code developed by \cite{takahashiFORMATIONOVERHEATEDREGIONS2016},
in which the zeroth and first moment equations of the radiation are solved with applying a M1 closure \citep{levermoreRelatingEddingtonFactors1984,kannoKineticSchemeSolving2013,sadowskiSemiimplicitSchemeTreating2013}.
In the following, the Greek suffixes indicate spacetime components, and the Latin suffixes indicate space components.
Length and time are normalized by gravitational radius ($r_{\rm g}=GM/c^2$) 
and its light crossing time ($t_{\rm g} =r_{\rm g}/c$), 
where $G$ is the gravitational constant and 
$M$ is the black hole mass.
Hereafter $c$ and $G$ are taken as unity.

The mass conservation equation is given by
\begin{eqnarray}
    \left( \rho u^\nu \right)_{;\nu} = 0, \label{eq:mas_con}
\end{eqnarray}
where $\rho$ is the proper mass density and $u^\mu$ is the fluid four velocity.
The energy momentum conservation equations for the magnetofluid 
and the radiation are given by
\begin{eqnarray}
    \left( {T_{\mu}}^{\nu} + {M_{\mu}}^{\nu} \right)_{;\nu} = G_\mu, \label{eq:ene_con}
\end{eqnarray}
and 
\begin{eqnarray}
    ({R_{\mu}}^{\nu})_{;\nu} = -G_{\mu}.
\end{eqnarray}
Here, the energy momentum tensor for fluid $T^{\mu\nu}$ is given by
\begin{eqnarray}
    {T_{\mu}}^{\nu} = (\rho + e + p_{\rm g})u_\mu u^\nu + p_{\rm g} \delta_{\mu}^{\nu},
\end{eqnarray}
where $e$ is the gas internal energy, $p_{\rm g}$ is the gas pressure, $\delta_{\mu}^{\nu}$ is the Kronecker delta.
The gas internal energy is related to the gas pressure by $e=(\Gamma -1)^{-1}p_{\rm g}$ with $\Gamma=5/3$ being the specific heat ratio.
The energy momentum tensor for electromagnetic field $M^{\mu\nu}$ is given by
\begin{eqnarray}
    {M_{\mu}}^{\nu} = 2p_{\rm m}u_\mu u^\nu + p_{\rm m}\delta_{\mu}^{\nu} -b_\mu b^\nu,
\end{eqnarray}
where $b^\mu$ is the magnetic four vector and $p_{\rm m}=b_\mu b^\mu/2$ is the magnetic pressure.
The magnetic four vector $b^\mu$ is related to its three vectors $B^i$ through
\begin{eqnarray}
    b^t &=& B^i u^\mu g_{i \mu},\\
    b^i &=& \frac{B^i + b^t u^i}{u^t},
\end{eqnarray}
where $g^{\mu\nu}$ is the metric.
The radiation energy momentum tensor is given by
\begin{eqnarray}
    R^{\mu \nu}=p_{\rm rad}
                \left( 
                    4u_{\rm rad}^{\mu}u_{\rm rad}^{\nu}+g^{\mu \nu}
                \right) , \label{eq:Rmn}
\end{eqnarray}
where $p_{\rm rad}$ is the radiation pressure and $u_{\rm rad}^{\mu}$ is the radiation frame’s four velocity.
The radiation four force $G^\mu$ is given by
\begin{eqnarray}
    G^\mu = &-&\rho \kappa_{\rm abs}({R_{\alpha}}^{\mu}u^\alpha 
                + 4\pi B u^\mu) \nonumber \\
            &-&\rho \kappa_{\rm sca}({R_{\alpha}}^{\mu}u^\alpha 
                + {R_{\beta}}^{\alpha}u_\alpha u^\beta u^\mu) 
                + {G^\mu}_{\rm comp}, \label{eq:Gmu}
\end{eqnarray}
where $\kappa_{\rm abs}$ and $\kappa_{\rm sca}$ are the opacities for free-free absorption and Thomson-scattering,
\begin{eqnarray}
    \kappa_{\rm abs} &=& 6.4\times10^{22}\rho T_{\rm g}^{-3.5} \ {\rm cm}^2 {\rm g}^{-1},  \\
    \kappa_{\rm sca} &=& 0.4 \ {\rm cm}^2 {\rm g}^{-1},
\end{eqnarray}
where $T_{\rm gas}$ is the gas temperature, which is related to the gas pressure as
\begin{eqnarray}
    p_{\rm g} = \frac{\rho k_{\rm B} T_{\rm g}}{\mu m_{\rm p}}.
\end{eqnarray}
Here $k_{\rm B}$ is the Boltzmann constant, $\mu=0.5$ is the mean molecular weight, and $m_{\rm p}$ is the proton mass.
The blackbody intensity is given by $B=a_{\rm rad}T_{\rm gas}^4/4\pi$ with $a_{\rm rad}$ being the radiation constant.
We included the thermal Comptonization as follows:
\begin{eqnarray}
    {G^\mu}_{\rm comp} = -\rho \kappa_{\rm sca} \hat{E}_{\rm rad} \frac{4 k_{\rm B}(T_{\rm e}-T_{\rm rad}) }{m_{\rm e}}u^\mu,
\end{eqnarray}
where $T_{\rm e}$ is the electron temperature, $\hat{E}_{\rm rad}$ is the comoving frame radiation energy density, $T_{\rm rad}=(\hat{E}_{\rm rad}/a_{\rm rad})^{1/4}$ is the radiation temperature, and $m_{\rm e}$ is the electron rest mass \citep{sadowskiPhotonconservingComptonizationSimulations2015a}.
We take $T_{\rm e}=T_{\rm g}$ for simplicity.
The induction equation is given by
\begin{eqnarray}
    \partial_t \left( \sqrt{-g}B^i \right)
   +\partial_j\left[\sqrt{-g}
                \left( b^i u^j -b^j u^i \right)
              \right] =0,
\end{eqnarray}
where $g={\rm det}(g_{\mu \nu})$.
We set $\Gamma =5/3$ for a gas, and $4/3$ for a radiation (see Equation (\ref{eq:Rmn})). The specific heat ratio of the mixed fluid is determined by their temperature. The assumption of $\Gamma=5/3$ could be violated for the relativistically hot gas. We find that the $T_{\rm g}$ is less than $m_{\rm p} c^2/k_{\rm B}$ in the whole simulation box after simulation. 
We believe the $\Gamma=5/3$ for the gas is a good approximation, even though more realistic equation of state is needed to implement for a detailed analysis \citep{mignoneEquationStateRelativistic2007,chattopadhyayEFFECTSFLUIDCOMPOSITION2009,sadowskiRadiativeTwotemperatureSimulations2017a,dihingiaShocksRelativisticViscous2019}.

We solve these equations in polar coordinate $(t,\ r,\ \theta,\ \phi)$ with the Kerr–Schild metric 
by assuming axisymmetry with respect to the rotation axis ($\theta=0$ and $\pi$).
The computational domain consists of $r=[R_{\rm in},R_{\rm out}]=[r_{\rm H},250r_{\rm g}]$, $\theta=[0, \pi]$,
where $r_{\rm H}=1+[1 + (a^*)^2]^{1/2}$ is a horizon radius 
with $a^*$ being the normalized spin parameter of black hole.
Number of numerical grid points is $(N_r,\ N_\theta,\ N_\phi)=(264,264,1)$.
A grid spacing exponentially increases with radius in the radial direction.
The radius of the $j$-th radial grid is given as $r(j)=\exp \left[ \ln r_{\rm H}+j \ln (R_{\rm out}/R_{\rm in})/N_r \right]$.
A polar angle of $k$-th polar grid is given by $\theta(k)=k\pi /N_{\theta}   + 0.25 \sin{[2\pi k/N_{\theta}]}$.
We adopted the outgoing boundary at inner and outer radius, and the reflective boundary is adopted at $\theta=0$ and $ \pi$.

We start simulations from an equilibrium torus 
\citep{fishboneRelativisticFluidDisks1976}
in which weak poloidal magnetic fields are initially embedded.
In the torus, the magnetic flux vector $A_\phi$ is proportional to $\rho$
and the plasma beta $\beta=(p_{\rm g}+p_{\rm rad})/p_{\rm m}$ is set to be 100 at most.
The inner edge of the initial torus is situated at $r=20r_{g}$, 
while its pressure maximum is situated at $r=33r_{g}$. 
We take $\rho_0 =1.4\times10^{-2}\ {\rm g\ cm^{-3}}$ as maximum mass density.

We employ $M=10M_\odot$ throughout the present study, where $M_\odot$ is the solar mass.
In total, we calculate nine models with $a^*=0,\ \pm 0.3,\ \pm 0.5,\ \pm 0.7,\ \pm 0.9$
until $t=5000t_{\rm g}$.
We note that the difference in the structure of the initial torus 
due to differences in $a^*$ is negligibly small.

\begin{figure}
\begin{center}
\includegraphics[scale=0.647]{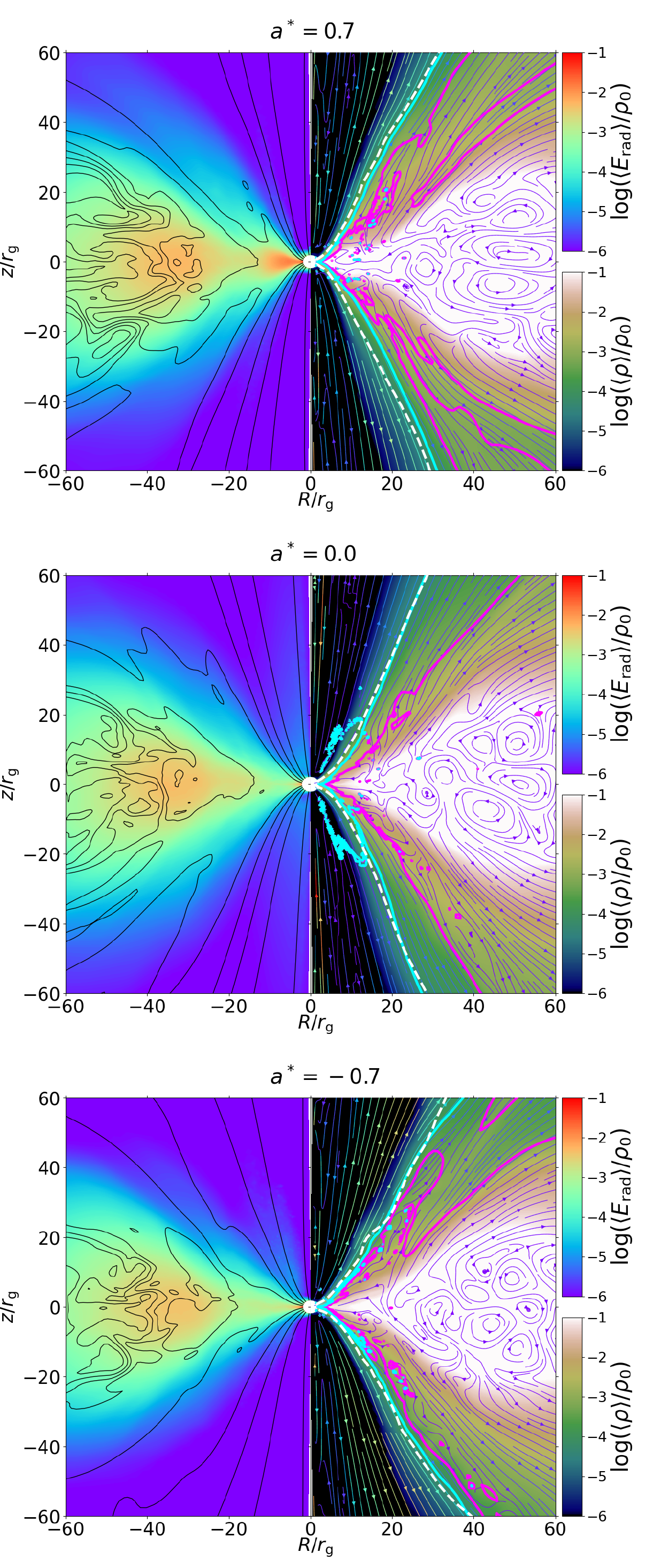}
\end{center}
\caption{
The structure of time-averaged ($t=3000 - 5000t_{\rm g}$) accretion disks, jets, and winds for $a^*=0.7$ (top), $a^*=0$ (middle), $a^*=-0.7$ (bottom).
In each panel, the left half shows the radiative energy density (colors) and the magnetic field lines (contours), and the right half shows the mass density (colors), $B_{e}=0$(magenta solid contours), $B_{e}=0.05$ (cyan solid contours), the photosphere (white dashed contours) and the mass flux direction (stream lines).
\label{fig:fig_ro_er}}
\end{figure}
%
\section{Results}\label{sec:res}
\subsection{Overview of Simulations}\label{sbsec:over}
\begin{figure}
\begin{center}
\includegraphics[scale=0.4]{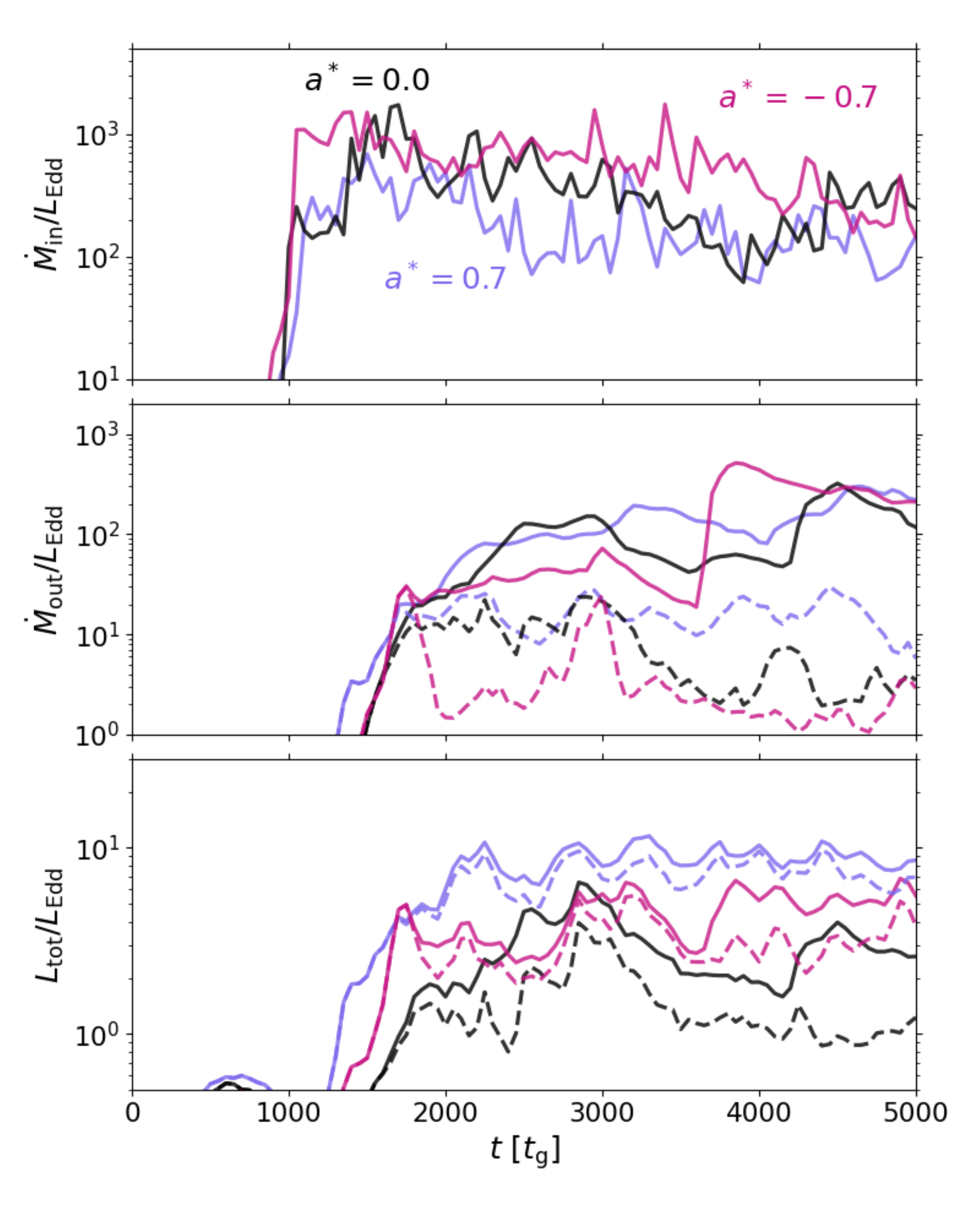}
\end{center}
\caption{
Time evolution of the mass accretion rate, the mass outflow rate and the total luminosity. Pink, black, and violet lines show results of $a^*=-0.7,\ 0$, and $ 0.7$, respectively.
The solid line shows the result of integration in the region of $B_e \geq 0$, and the dashed line shows the result in the region of $B_e \geq 0.05$.
\label{fig:fig_timedep}}
\end{figure}
At the beginning, we introduce in Figure {\ref{fig:fig_ro_er}} the global structure of 
supercritical accretion flows, jets, and winds (we will define later) for the case with $a^*=0.7$ (top), $0$ (middle), and $-0.7$ (bottom).
The color contour on the left side represents the radiative energy density, and the solid lines indicate the magnetic field lines.
The radiation energy density is expressed as
\begin{eqnarray}
    E_{\rm rad}=n_\alpha n_\beta R^{\alpha \beta},
\end{eqnarray}
where $n_\alpha = (-\alpha,0)$ is the normal observer's four velosity with $\alpha=(-g^{t t})^{-1/2}$ being the lapse function.
The mass density is shown by the color contour on the right side.
The streamlines,
which are calculated from the velocity field in the zero angular momentum observer (ZAMO) frame,
are overlaid with the arrows.
In this figure, all physical quantities are time-averaged over $t=3000-5000t_{\rm g}$.
Here we note that the flow is basically quasi-steady during this time-averaged period (see below), but there are some fluctuations. In particular, somewhat larger fluctuations appear in the region where the mass density is much less than $10^{-2} \rho_0$ (where the gas flows outward).
Hereafter, the time averaged quantities over $t=3000-5000t_{\rm g}$ is denoted with $\langle\ \rangle$.

The white dashed lines show the photosphere ($\tau_{\rm tot}=1$), where $\tau_{\rm tot}$ is the total optical depth measured from the polar axis,
\begin{equation}
    \tau_{\rm tot}=\int \gamma \rho(\kappa_{\rm abs}+\kappa_{\rm sca})\sqrt{g_{\theta \theta}}d \theta,
\end{equation}
\citep{mckinneyThreedimensionalGeneralRelativistic2014}.
Here, $\gamma=\alpha u^t$ is the Lorentz factor.
The magenta and cyan solid lines indicate the surface where Bernoulli parameter, $B_e$, is $0$ and $0.05$, respectively. Here $B_e$ is given by
\begin{equation}
    B_e \equiv-\frac{{T_t}^r+{M_t}^r+{R_t}^r+\rho u^r}{\rho u^r},
\end{equation}
\citep{sadowskiEnergyMomentumMass2013c,sadowskiPowerfulRadiativeJets2015a,sadowskiThreedimensionalSimulationsSupercritical2016a}.
Note that $B_e \ge 0$ is the condition for the steady flow to reach infinity, 
and the flows with $B_e \ge 0.05$ can have $\sim 30\%$ of the speed of light at infinity.

In all models, we distinguish the whole structure into three regions. 
One of them is the disk region with $B_e<0$,
and another is the jet region with $B_e \ge 0.05$ around the rotation axis.
The remaining one is the wind region, which appears between these two regions ($0\le B_e<0.05$). 
We must note that the boundary between the jet and the wind is not well defined, but they are smoothly connected. Thus, the jet boundary defined as $B_e \ge 0.05$ might not agree with the observed one. 
Despite this, we think the definition of the jet boundary by $B_e \ge 0.05$ is useful to distinguish the fast jet from the slower wind. 
In the disk region, we can see that the gas flows toward the black hole with the turbulent motion.
The large number of photons is accumulated since the disk is very optically thick.
In the jet region, the rarefied matter is blown away
in all models except very vicinity of the rotation axis when $a^*=0$.
The radial velocity, which is measured in the ZAMO frame, in that region reaches a maximum of $\sim 60\%$ of the speed of light
for the case with $a^*=0.7$ and $a^*=-0.7$, 
although it is less than $\sim 30\%$ of the speed of light when $a^*=0$.
The magnetic field lines in the jet region have a radial structure along the gas flow
although the complex structure arises in the disk region.
In the wind region, where the Bernoulli parameter is roughly $0 \le B_e < 0.05 $,
relatively dense matter ($\rho/\rho_0 \sim 10^{-5}-10^{-2}$) 
moves outward at $\sim 10-20\%$ of the speed of light, regardless of $a^*$.
Although not shown in Figure {\ref{fig:fig_ro_er}}, accretion disk, winds, and jets also appear in all other models.

In the present study, 
only the model with $a^*=0.9$ eventually reaches MAD limit
\citep{tchekhovskoyProgradeRetrogradeBlack2012a}
and shows violent behavior such as a rapid increase or decrease in accretion rate and outflow rate. 
Therefore, we mainly compare the models with $a^*=\pm 0.7$ with the model with $a^*=0$.

\subsection{Mass Flow and Luminosity}\label{sbsec:MandL}

Next, we measure the mass accretion rate at the event horizon, as
\begin{eqnarray}
    \dot{M}_{\rm in} = -2\pi \int_0^\pi \rho u^r \sqrt{-g}\ d\theta.
\end{eqnarray}
The mass outflow rate is calculated at $r=200r_{\rm g}$ as 
\begin{eqnarray}
    \dot{M}_{\rm out} &=& 2\pi \int_0^{\theta_{B_e}} \rho u^r \sqrt{-g}\ d\theta  \nonumber \\
                     &\ &+2\pi \int_{\theta'_{B_e}}^\pi \rho u^r \sqrt{-g}\ d\theta,
\end{eqnarray}
where $\theta_{B_e}$ ($\theta'_{B_e}$) is the angle to identify the jet and wind region for $\theta<\pi/2$ ($\theta>\pi/2$). We set the angle of $B_e=0.05$, which indicates the boundary of jet and wind regions. We also set the angle of $B_e=0$, which separates the wind and disk regions.
The angle of $B_e=0.05$, which corresponds to the opening angle of the jets,
is slightly larger for the models with $a^*\pm 0.7$ than the model with $a^*=0$.
The jet opening angle is around $35^\circ (a^*=-0.7),\ 20^\circ (a^*=0)$, and $30^\circ(a^*=0.7)$ at $r=200r_{\rm g}$.
Furthermore, the opening angle of the jet roughly corresponds to the opening angle of the photosphere.
(see Figure {\ref{fig:fig_ro_er}})

The total luminosity is evaluated at $r=200r_{\rm g}$ by 
\begin{eqnarray}
   L_{\rm tot} &=& L_{\rm rad}+L_{\rm mag}+L_{\rm kin},
\end{eqnarray}
where 
$L_{\rm rad}$ is the radiative luminosity,
\begin{eqnarray}
    L_{\rm rad} &=& -2\pi \int_0^{\theta_{B_e}} {R_t}^r \sqrt{-g}\ d\theta  \nonumber \\
                 &\ &-2\pi \int_{\theta'_{B_e}}^\pi {R_t}^r \sqrt{-g}\ d\theta , \label{eq:Lrad}
\end{eqnarray}
$L_{\rm mag}$ is the Poynting power
(luminosity, which is calculated by Poynting flux),
\begin{eqnarray}
    L_{\rm mag} &=& -2\pi \int_0^{\theta_{B_e}} {M_t}^r \sqrt{-g}\ d\theta  \nonumber \\
                 &\ & -2\pi \int_{\theta'_{B_e}}^\pi {M_t}^r \sqrt{-g}\ d\theta , \label{eq:Lmag}
\end{eqnarray}
$L_{\rm kin}$ is the kinetic power
(luminosity, which is calculated by kinetic flux), 
at $r=200r_{\rm g}$,
\begin{eqnarray}
   L_{\rm kin} &=& -2\pi \int_0^{\theta_{B_e}} {K_t}^r \sqrt{-g}\ d\theta  \nonumber \\
                &\ & -2\pi \int_{\theta'_{B_e}}^\pi {K_t}^r \sqrt{-g}\ d\theta . \label{eq:Lkin}
\end{eqnarray}
Here $-{K_t}^r$ is the energy momentum tensor of the fluid minus 
the thermal energy, the rest mass energy, and the potential energy components, 
\begin{equation}
    -{K_t}^r = -\rho u^r\left( u_t + 
             \sqrt{-g_{tt}} 
             \right), \label{eq:kin}
\end{equation}
\citep{sadowskiEnergyFlowsThick2016},
and becomes the kinetic flux in the non-relativistic limit.

Figure {\ref{fig:fig_timedep}} shows the time evolution of the mass accretion rate (top), 
the mass outflow rate (middle), and the total luminosity (bottom) for $a^*=0,\ \pm0.7$
normalized by Eddington luminosity given by
\begin{eqnarray}
    L_{\rm Edd}=1.25\times10^{39} \left( \frac{M}{10M_{\odot}}\right) \ {\rm erg\ s^{-1}}.
\end{eqnarray}
As shown in the top panel,
the mass accretion rate rapidly increases at around $t=1000t_{\rm g}$ 
since the gas from the initial torus reaches the event horizon in all models.
At $t \ge 1000t_{\rm g}$, a quasi-steady supercritical disk is formed, 
and the accretion rate always exceeds the Eddington limit ($L_{\rm Edd}$).
The average mass accretion rate is $\sim 470$, $240$, and $150 L_{\rm Edd}$ for $a^*=-0.7,\ 0$, and $0.7$, respectively.
Although there are some fluctuations, the mass outflow rate remains above $L_{\rm Edd}$ in all models
at $t \geq 3000t_{\rm g}$, meaning that the powerful winds are launched.
It is also found that the outflow rate in the region of $B_e\ge0$ (solid line, jet+wind region) 
is much higher than that in the region of $B_e\ge0.05$ (dashed line, jet region).
Thus, we find that the matter is mainly blown passing through the wind region.

\begin{figure}
\begin{center}
\includegraphics[scale=0.41]{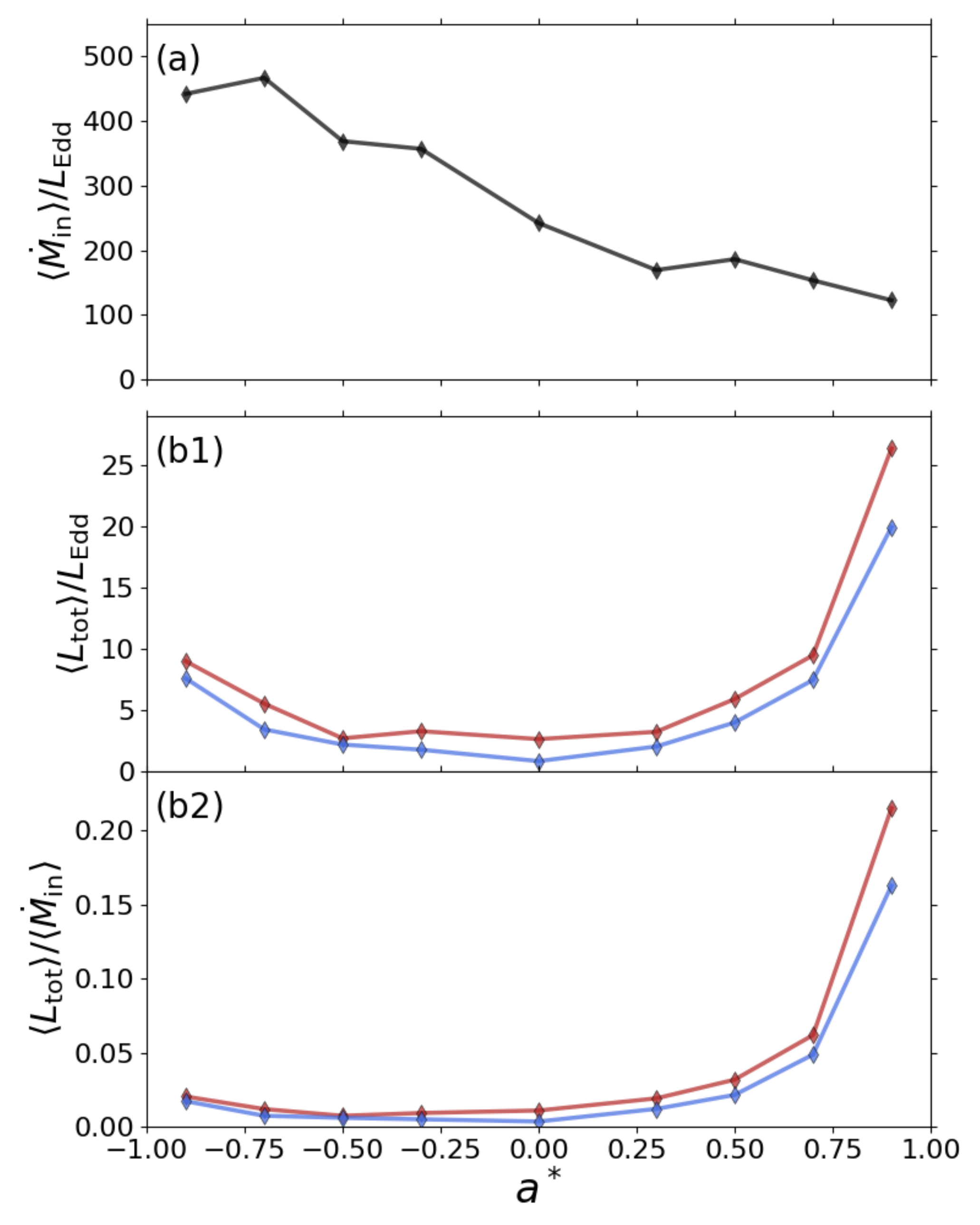}
\end{center}
\caption{
Black hole spin dependence of the mass accretion rate $\langle \dot{M}_{\rm in}\rangle$ (panel a), the total luminosity $\langle L_{\rm tot} \rangle$ (panel b1), the energy outflow efficiency $\langle L_{\rm tot} \rangle/\langle \dot{M}_{\rm in} \rangle$ (panel b2).
The mass accretion rate is measured at the event horizon, and the red (blue) line shows the integrated flux through the sphere at $r=200r_{\rm g}$, which satisfies the condition $B_e \geq 0$ ($B_e \geq 0.05$).
\label{fig:fig_MLspin}}
\end{figure}

We can see in the bottom panel that the total luminosity in the region of $B_e\ge0$ (solid line) 
is kept above $L_{\rm Edd}$ in all models at $t \geq 3000t_{\rm g}$.
This panel indicates that the total luminosity for $B_e\ge0.05$ (dashed line) 
is slightly smaller than that in the region of $B_e\ge0$ (solid line), but not by much.
Thus, we can conclude that the energy is mainly released through the jet region unlike the ejecting matter.
Here, the luminosity, accretion rate, and mass outflow rate are measured at $200r_{\rm g}$. 
We note that they are almost constant with radius far from the black hole for $ r>150r_{\rm g} $, and the mass accretion rate, along with outflow rate, reach steady states at $t \geq 3000r_{\rm g}$. Thus, these results do not depend on the choice of the radius for $r > 150 r_{\rm g}$. 
Hereafter, we take time average between $t=3000t_{\rm g}$ and $5000t_{\rm g}$ since we can recognize that the disk, wind, and jet reach a steady state.

\begin{figure}
\begin{center}
\includegraphics[scale=0.42]{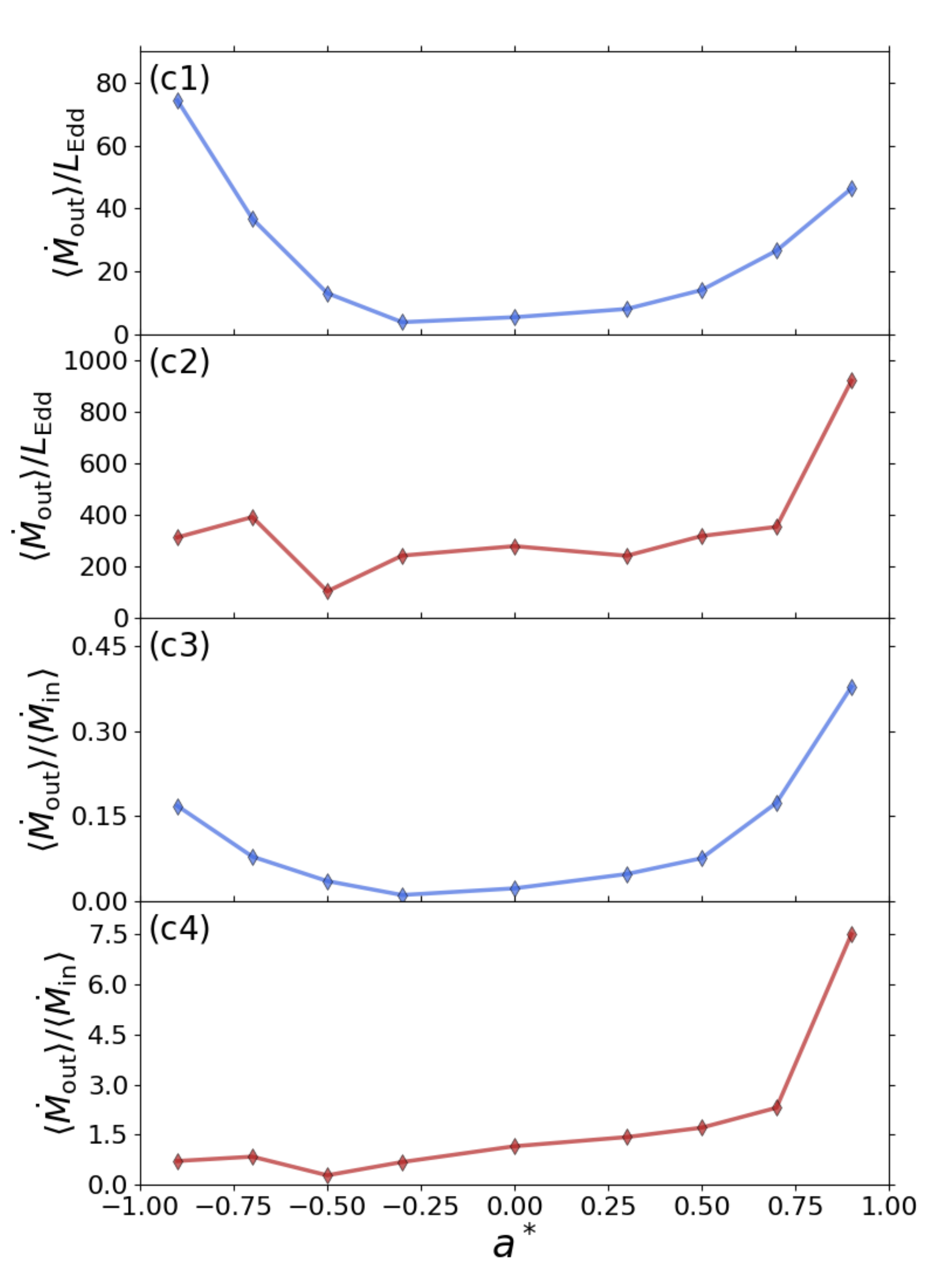}
\end{center}
\caption{
Black hole spin dependence of the mass outflow rate $\langle \dot{M}_{\rm out} \rangle$ (panel c1 and c2), and the mass outflow efficiency $\langle \dot{M}_{\rm out} \rangle/\langle \dot{M}_{\rm in} \rangle$ (panel c3 and c4).
The red (blue) line shows the integrated flux through the sphere at $r=200r_{\rm g}$, which satisfies the condition $B_e \geq 0$ ($B_e \geq 0.05$).
\label{fig:fig_Mspin}}
\end{figure}
\begin{figure*}
\plotone{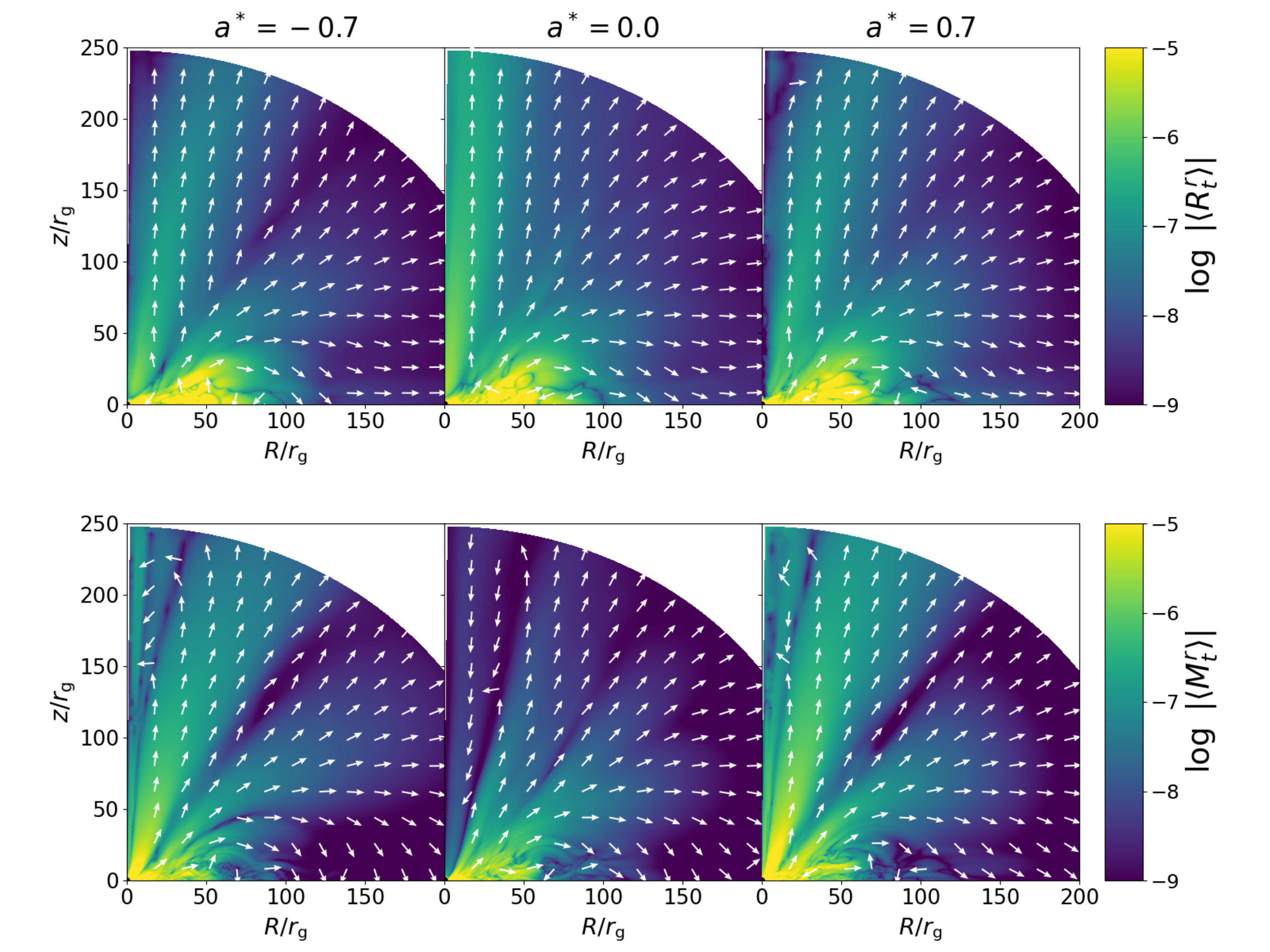}
\caption{
The time averaged radiative flux (top) and Poynting flux (bottom) for $a^*=-0.7$ (left), $a^*=0$ (middle), $a^*=0.7$ (right).
Colors show the magnitudes of fluxes, while arrows show their direction on the poloidal plane.
\label{fig:fig_flux}}
\end{figure*}

In Figure \ref{fig:fig_MLspin}, we plot $\langle \dot{M}_{\rm in} \rangle/L_{\rm Edd}$ (panel a), 
$\langle L_{\rm tot} \rangle /L_{\rm Edd}$ (panel b1),
$\langle L_{\rm tot} \rangle/\langle \dot{M}_{\rm in} \rangle$ (panel b2) 
as a function of $a^*$ from top to bottom.

The mass accretion rate tends to decrease with increasing $a^*$ (see panel a).
For instance, the mass accretion rate is $\sim 470 L_{\rm Edd}$ for $a^*=-0.7$, 
$\sim 240 L_{\rm Edd}$ for $a^*=0$, 
and $\sim 150 L_{\rm Edd}$ for $a^*=0.7$.

It is found in the panel b1 that 
the total luminosity tends to increase with $|a^*|$. 
We attribute this result to the effect of the Blandford-Znajek (BZ) mechanism \citep{blandfordElectromagneticExtractionEnergy1977} (we will discuss later).
Since the total luminosity is more sensitive to the spin parameter in the range of $a^*>0$,
$\langle L_{\rm tot} \rangle$ is higher for the prograde case than the retrograde case with same $|a^*|$.
The total luminosity in the region of $B_e \ge 0$ for $a^*=0.7$
is $\sim 9 L_{\rm Edd}$, that is higher than the luminosity for $a^*=0$ ($\sim 3 L_{\rm Edd}$)
and for $a^*=-0.7$ ($\sim 5 L_{\rm Edd}$).
As noted before, the opening angle of the jet is larger for a rapidly rotating case, but the size of the solid angle of the jet is only 1.5 times different between $a^*=0$ and $a^*=0.7$. Thus, the higher luminosity for a rapidly rotating black hole originates from the higher flux rather than the larger solid angle.
Note that the difference between the luminosity in the region of $B_e\ge0$ (red) and that in the region of $B_e\ge0.05$ (blue)
is very small, which means that the energy is mainly released through the jet region
as we have already mentioned in Figure \ref{fig:fig_timedep}.

Since the mass accretion rate tends to be higher for $a^*<0$ than for $a^*>0$,
the enhancement of the energy outflow efficiency ($\langle L_{\rm tot} \rangle/\langle \dot{M}_{\rm in} \rangle$) 
is more pronounced when $a^*$ is large (see panel b2).
The panel b2 shows that the energy outflow efficiency becomes
$\gtrsim 0.05$ only when $a^* \geq 0.7$.

Figure \ref{fig:fig_Mspin} shows the mass outflow rate $\langle \dot{M}_{\rm out} \rangle$ (panel c1 and c2), and the mass outflow efficiency $\langle \dot{M}_{\rm out} \rangle/\langle \dot{M}_{\rm in} \rangle$ (panel c3 and c4). The outflow rate is measured where $B_e \geq 0.05$ (panel c1 and c3), and where $B_e \geq 0$ (c2 and c4).
As we have mentioned in Figure \ref{fig:fig_timedep}, 
the mass is mainly ejected through the wind region, so that the outflow rate for $B_e \geq 0.05$ (blue line) 
is much lower than that for $B_e \geq 0$ (red line)
(see panels c1, c2, c3, and c4).
It is found that $\langle \dot{M}_{\rm out} \rangle$ 
and $\langle \dot{M}_{\rm out} \rangle/\langle \dot{M}_{\rm in} \rangle$ for $B_e \geq 0.05$ 
tend to increase with increasing $|a^*|$. 
Such a trend may be related to the BZ effect.
On the other hand, 
although the mass outflow rate in the region of $B_e \geq 0$ is 
very high when $a^*=0.9$,
there is no clear dependence of $\langle \dot{M}_{\rm out} \rangle$ on $a^*$ in panel c2.
The ratio of $\langle \dot{M}_{\rm out} \rangle$ to $\langle \dot{M}_{\rm in} \rangle$ 
slightly increases with $a^*$ (see panel c4).
The mechanism of gas ejection from the surface of the accretion disk 
is mainly radiative force, but magnetic force also comes into play when $|a^*|$ is large (we will discuss in Section \ref{sbsec:Mdot_out}).

\begin{figure}
\begin{center}
\includegraphics[scale=0.4]{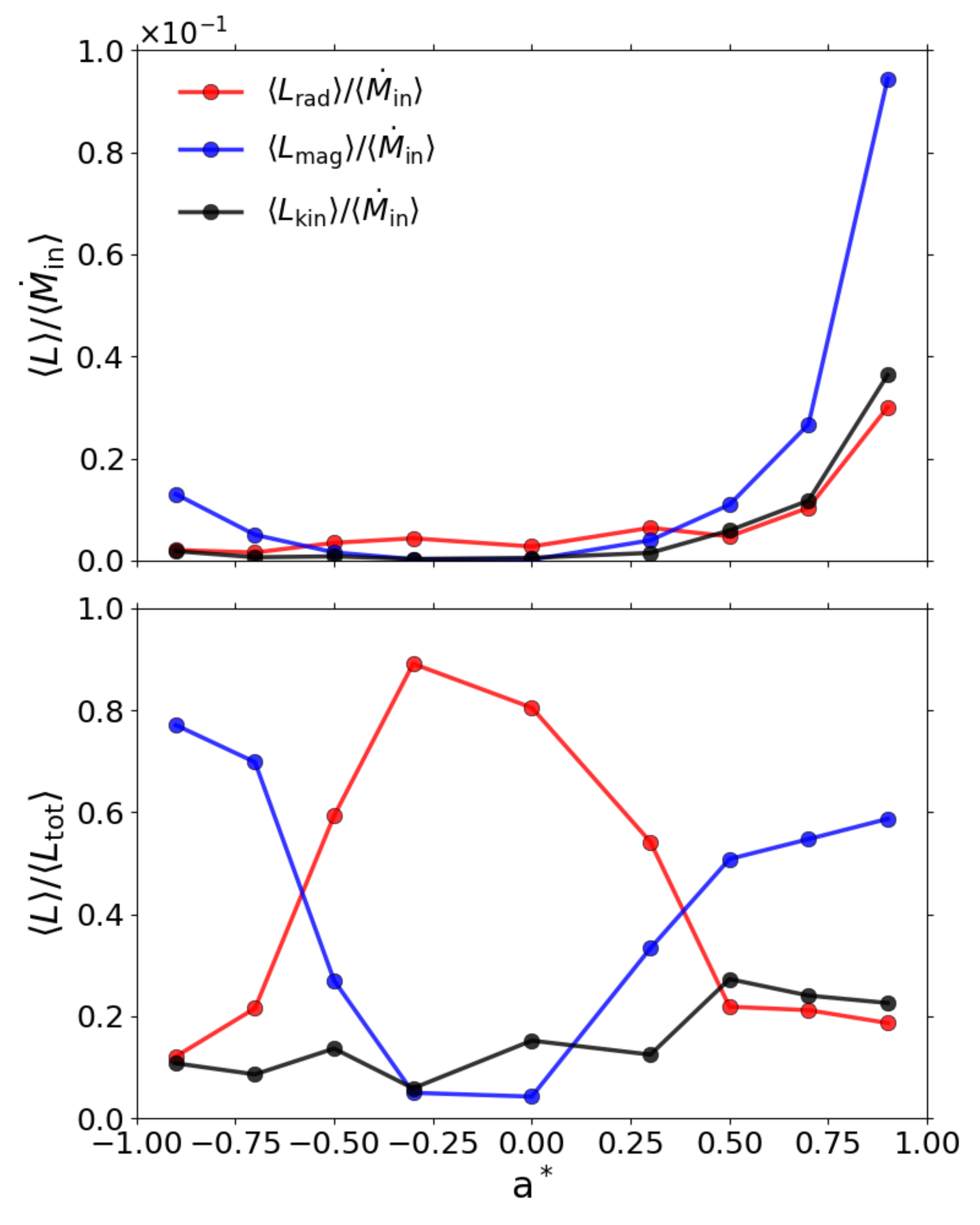}
\end{center}
\caption{
Black hole spin dependence for each energy outflow efficiency (top) and ratio of each luminosity component to the total luminosity (bottom).
Radiative component (red), magnetic component (blue) and kinetic component (black).
The integration region is $B_e \geq 0.05$ at $r=200r_\mathrm{g}$.
\label{fig:fig_Lspin}}
\end{figure}

The top panel in Figure \ref{fig:fig_flux} represents the radiative flux.
Color contour shows the magnitude of the radial component of the radiative flux $|{R_t}^r|$ 
and arrows denotes the direction of the radiative flux on the poloidal plane.
We find that the magnitude and direction of the radiative flux do not depend much on the black hole spin.
In all models, the radiative flux near the equatorial plane is the highest, 
and the radiative flux is also strong in the jet region near the rotation axis.
The direction of the radiative flux is basically outward, but in the disk, 
there are places where the flux turns inward or toward the equatorial plane.
This is thought to be due to the turbulent motion of the optically thick matter within the disk.

In the bottom three panels of Figure \ref{fig:fig_flux}, we show the magnitude of the radial component of the Poynting flux $|{M_t}^r|$ by color contour
and the direction of the Poynting flux by arrows.
As with $|{R_t}^r|$,  $|{M_t}^r|$ is higher in the disk region in all models.
This is thought to be due to the amplification of the magnetic field inside the disk.
In addition, the enhanced outward Poynting flux appears in the regions of jet and wind 
extending diagonally upward from the black hole in models with $a^*=\pm 0.7$, 
except very vicinity of the rotation axis.
Although the outward Poynting flux in the wind region is slightly enhanced in the model with $a^*=0$,
the magnitude of it is much smaller than the other models.
Therefore, the enhanced Poynting fluxes in the jet and wind regions that appear 
in models of rotating black holes may be due to the BZ mechanism.
Indeed, we can confirm that 
the resulting Poynting flux at the event horizon, $-M_t^r (r=r_{\rm H})$,
is consistent with the BZ flux 
$F_{\rm BZ} |_{r=r_{\rm H}}$
derived 
in \citet{mckinneyMeasurementElectromagneticLuminosity2004a},
\begin{eqnarray}
   F_{\rm BZ}|_{r=r_{\rm H}} = 2(B^r)^2 \omega r_{\rm H} ( \Omega_{\rm H} -\omega ) \sin^2{\theta}.
\end{eqnarray}
Here $\Omega_{\rm H} \equiv a^*/2r_{\rm H}$ and $\omega$ 
are the rotation frequency of the black hole and the magnetic field at the event horizon.
For example, the magnitudes calculated at $\theta=\pi/4$ are $\sim 6\times10^{-4}\rho_0$ ($-M_t^r$) and $\sim 4\times10^{-4}\rho_0$ ($F_{\rm BZ}$) for $a^* \pm 0.7$. 
We note that direction of black hole spin, accretion flow, and magnetic field rotation are aligned even for $a^*<0$ at the event horizon.

It is clearly understood in Figure \ref{fig:fig_Lspin} that the BZ effect enhances the Poynting flux.
The top panel in the figure shows the energy outflow efficiency 
via the radiative flux, $\langle L_{\rm rad} \rangle/\langle \dot{M}_{\rm in} \rangle$,
the Poynting flux, $\langle L_{\rm mag} \rangle/\langle \dot{M}_{\rm in} \rangle$, 
and the kinetic flux, $\langle L_{\rm kin} \rangle/\langle \dot{M}_{\rm in} \rangle$,
as a function of $a^*$.
In the bottom panel,
we also plot $\langle L_{\rm rad} \rangle$, $\langle L_{\rm mag} \rangle$, 
and $\langle L_{\rm kin} \rangle$ normalized by $\langle L_{\rm tot} \rangle$.
Here, $L_{\rm rad}$, $L_{\rm mag}$, and $L_{\rm kin}$ in the figure 
indicates the luminosities in the jet regions (see Equations (\ref{eq:Lrad})-(\ref{eq:Lkin})).

The luminosity due to the Poynting flux tends to increase with an increase of $|a^*|$.
Although $\langle L_{\rm mag} \rangle/\langle \dot{M}_{\rm in} \rangle$ is 
only $\sim10^{-4}$ in the model with $a^*=0$,
it exceeds $\sim0.01$ in the case of $a^*=-0.9$ and $a^* \geq 0.5$.
The luminosity calculated from the Poynting flux 
is responsible for more than 50\% of the total luminosity
for $a^* \geq 0.5$ and $a^* \leq -0.7$ (see bottom panel).

The energy outflow efficiency via the radiative flux exhibits a slightly increasing trend with increasing $a^*$,
but is not as sensitive as that via the Poynting flux.
Although $\langle L_{\rm rad} \rangle$ is higher than $\langle L_{\rm mag} \rangle$ for $a^* \sim 0$, 
we find $\langle L_{\rm rad} \rangle < \langle L_{\rm mag} \rangle$ for 
the case of $a^* \geq 0.5$ and $a^* \leq -0.7$.
As a result, the radiative luminosity becomes dominant in the range of $a^*=-0.5$ and $a^*=0.3$
as shown in the bottom panel.
Also, the kinetic power is enhanced in the models with large $a^*$.
However, $\langle L_{\rm kin} \rangle$ is never the most dominant at $r=200r_\mathrm{g}$

Our results suggest that 
the black hole spin enhances the energy flux, especially the Poynting flux. 
The cause is probably the BZ effect.
Supercritical flows release the energy mainly by radiation for slowly rotating black hole, 
and by Poynting flux for rapidly rotating black hole.
\subsection{Mass Outflow Rate}\label{sbsec:Mdot_out}
%
\begin{figure*}
\begin{center}
\includegraphics[scale=0.55]{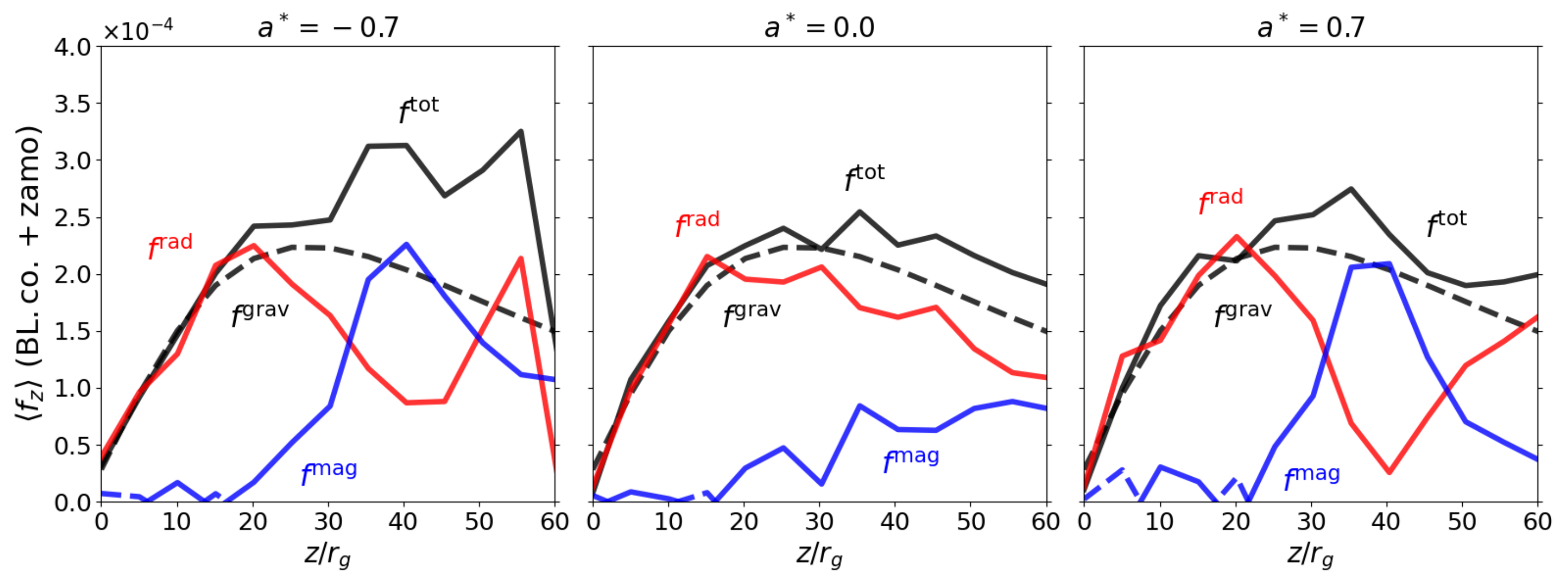}
\end{center}
\caption{
The time averaged force per unit mass in $z(=r\cos{\theta})$ direction as a function of $z$ for $a^*=-0.7$ (left), $a^*=0$ (middle), $a^*=0.7$ (right).
Red, blue, and black lines show the radiative force, the magnetic force, and the total force of radiation plus magnetic force.
Black dashed lines show gravitational force, and dashed lines indicate force in the negative direction.
The radius at which these are measured is $R(=r\sin{\theta})=50r_{\rm g}$.
Forces are measured in Boyer-Lindquist coordinates (BL. co.) and in the ZAMO frame.
\label{fig:fig_force}}
\end{figure*}
In the present simulations, 
the matter ejected from the disk surface is mainly blown away through the wind region
as we have mentioned in Section \ref{sbsec:MandL}.
Here, we discuss about forces, which launch the matter from the disk.

The general relativistic force per unit mass, 
which is modified to more precisely describe the forces related to the magnetic field in the previous study \citep{mollerAccelerationWindOptically2015a}, 
is shown below.
Assuming a steady state, 
the equation of motion is obtained from the energy momentum conservation for magnetofluids as follows:
\begin{eqnarray}                            
    && \frac{\rho}{w}u^i \partial_i u_\nu + 
                            \left(
                                1-\frac{\rho}{w}
                            \right)
                            \partial_\nu (u^i u_\nu) \nonumber \\
                         & = & - \frac{1}{w}\partial_\nu p_{\rm g} 
                            -  \frac{1}{w}\partial_\nu p_{\rm m} 
                            +  \frac{1}{w}\partial_i (b_\nu b^i)
                            +  \frac{1}{w} G_\nu \nonumber \\
                           && - \frac{1}{w}u_\nu u^i \partial_i (w-\rho) 
                            + \frac{1}{w} \left( {T_\lambda}^\kappa + {M_\lambda}^\kappa \right) \tensor{\Gamma}{^\lambda_\nu_\kappa} \nonumber \\
                           && -  \frac{ ({T_\nu}^i+{M_\nu}^i) -\rho u_\nu u^i }{w}
                               \frac{\partial_i (\sqrt{-g})}{\sqrt{-g}}, 
                               \ \label{eq:fc3}
\end{eqnarray}
where 
$w=\rho + e + p_{\rm g}+2p_{\rm m}$ denotes the relativistic enthalpy, 
and $\tensor{\Gamma}{^\lambda_\mu_\nu}$ is the christoffel symbol. 
Based on this equation, 
the gas pressure gradient force, the radiative force, the magnetic force, and the metric force
are defined by
\begin{eqnarray}
    f_\nu^{\rm gas} &=& -\frac{1}{w}\partial_\nu p_{\rm g}, \label{eq:fgas}\\
    f_\nu^{\rm rad} &=& \frac{1}{w} G_\nu, \\
    f_\nu^{\rm mag} &=& -\frac{1}{w}\partial_\nu p_{\rm m}
                        +\frac{1}{w}\partial_i (b_\nu b^i) \nonumber \\
                    &\ &
                    -\frac{1}{w}u_\nu u^i \partial_i (w-\rho)\nonumber \\
                    &\ &
                    +\frac{1}{w}{{M_\lambda}^\kappa} \tensor{\Gamma}{^\lambda_\nu_\kappa}
                            - \frac{{{M_\nu}^i} -\rho u_\nu u^i }{w}
                            \frac{\partial_i (\sqrt{-g})}{\sqrt{-g}},   \\
    f_\nu^{\rm metric} &=& \frac{1}{w}{{T_\lambda}^\kappa} \tensor{\Gamma}{^\lambda_\nu_\kappa}
                            - \frac{{{T_\nu}^i} -\rho u_\nu u^i }{w}
                            \frac{\partial_i (\sqrt{-g})}{\sqrt{-g}}. \label{eq:fmet}
\end{eqnarray}
All terms related to the magnetic field are included in the magnetic force, $f_\nu^{\rm mag}$.
We use the gravity in the non-relativistic limit,
\begin{eqnarray}
    f_r^{\rm grav} =& -\frac{1}{r^2}, \label{eq:fgrv} 
\end{eqnarray}
since we don't investigate the very vicinity of the black hole in this section.

Figure {\ref{fig:fig_force}} 
shows the component of the force perpendicular to the equatorial plane as a function of $z(=r\cos{\theta})$
at $R(=r\sin{\theta})=50r_{\rm g}$.
Here, these forces are transformed to the ZAMO frame in Boyer-Lindquist coordinates.
At the deep inside the disk ($z \lesssim 20 r_{\rm g}$), 
the radiative force is more dominant than the magnetic force and is balanced with gravity in all models.
Around the disk surface ($z \sim 20-30 r_{\rm g}$ and $B_e \sim 0$), the radiative force is stronger than the magnetic force, 
and the sum of the radiative and magnetic forces ($f^{\rm tot}$) slightly exceeds gravity.
Thus, the gas is launched from the disk surface . 
Here we note that 
the gas pressure gradient force and metric force are negligibly small 
compared to radiative force or magnetic force.

Since $f^{\rm tot}$ is stronger than the gravity force even at the region above the disk ($z \gtrsim 30 r_{\rm g}$),
the gas is accelerated.
However, whether the radiative force or the magnetic force is stronger depends on the black hole spin.
The magnetic force steeply increases from inside the disk $z<20r_\mathrm{g}$ to wind region, and exceeds the radiative force at around $z \gtrsim 35 r_{\rm g}$
for the models with $a^*=\pm 0.7$.

\section{Discussion}\label{sec:dis}
\subsection{Astrophysical Implications}
\begin{figure}
\begin{center}
\includegraphics[scale=0.42]{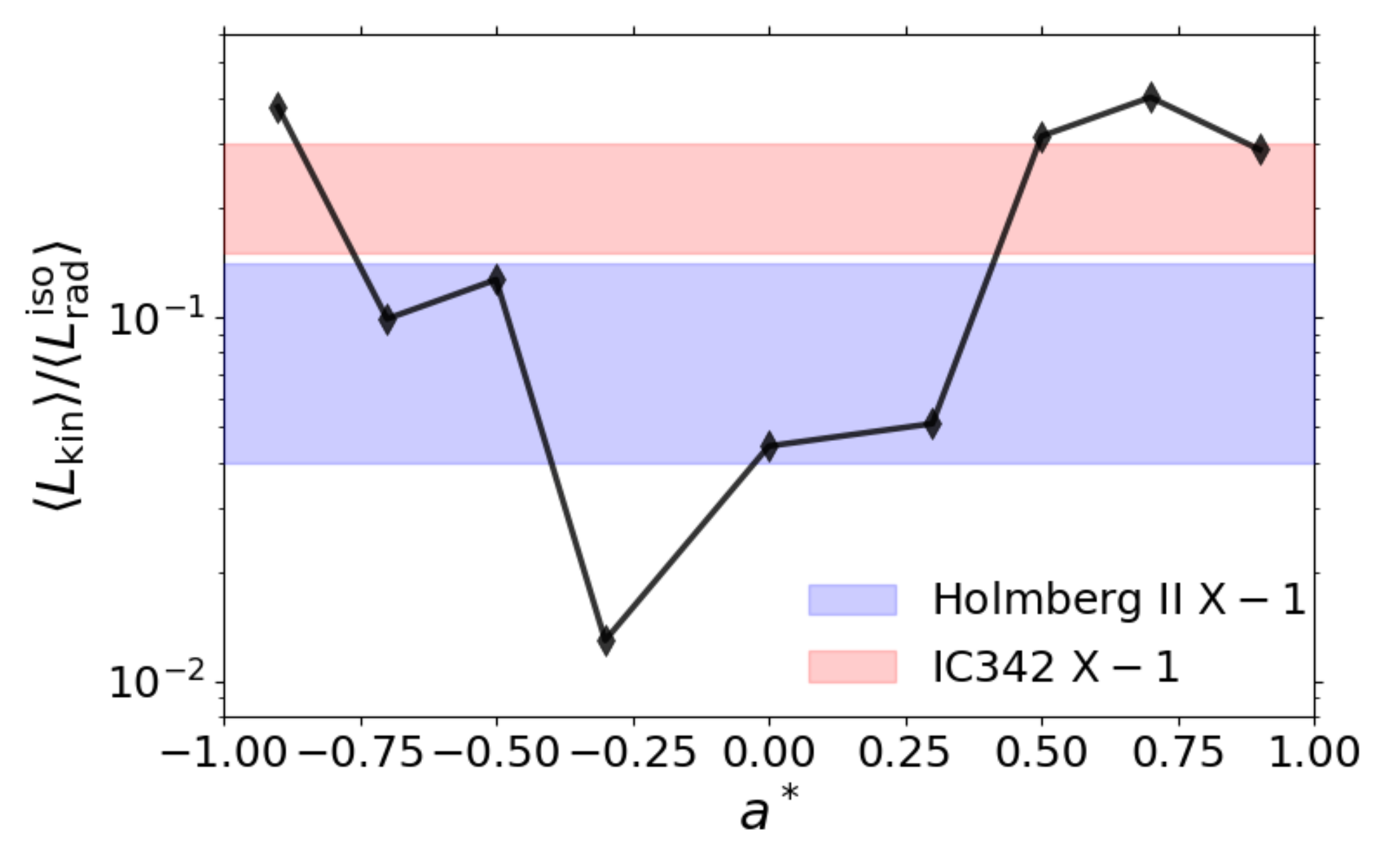}
\end{center}
\caption{
Black hole spin dependence of the power ratio of the kinetic and isotropic radiative Luminosity $L_{\rm kin}/L_{\rm rad}^{\rm iso}$ (black line). 
The blue (red) regions are the power ratios of ULX Holmberg II X-1 (IC342 X-1) suggested by the observations.
The calculation region is $B_e \ge 0.05$ at $r = 200r_{\rm g}$.
\label{fig:fig_obs}}
\end{figure}

Now, we compare our results with ULX observations. Figure \ref{fig:fig_obs} shows the spin dependence of isotropic X-ray luminosity $L_{\rm rad}^{\rm iso}$, where
\begin{eqnarray}
   L_{\rm rad}^{\rm iso} &=& 2\pi \Biggl[
                            -\frac{ \int_0^{\theta_{B_e}} {R_t}^r \sqrt{-g}\ d \theta}
                            { \int_0^{\theta_{B_e}} \sin{\theta} \ d \theta}  \nonumber \\
                         &\ &\ \ \ - \frac{ \int_{\theta'_{B_e}}^\pi {R_t}^r \sqrt{-g}\ d \theta}
                                        {\int_{\theta'_{B_e}}^\pi \sin{\theta} \ d \theta}
                                  \Biggr ].
   \end{eqnarray}
As \citet{kitakiOutflowSuperEddingtonFlow2021a} noted, the ratio of kinetic luminosity $L_{\rm kin}$ and isotropic X-ray luminosity is $\sim 0.04-0.14$ for ULX Holmberg II X-1, and $\sim 0.15-0.3$ for ULX IC342 X-1 \citep{kaaretHighresolutionImagingHe2004,abolmasovSpectroscopyOpticalCounterparts2007,shidatsuNuSTARSwiftObservations2017}. These values are specified with red and blue hatches in Figure \ref{fig:fig_obs}. Even though model $a^*=-0.3$ is an exception, the value of ULX Holmberg II X-1 is consistent with our models of between $a^*=-0.7$ and $0.3$. Also, there is not an inconsistency between ULX IC342 X-1 and our models of $a^* \ge 0.5$ and $a^* = -0.9$. These results indicate that the Holmberg X II X-1 originates from a slowly, or mildly rotating black hole, and IC342 X-1 contains relatively a rapidly rotating black hole.
Here, we note that the ratio of $L_{\rm kin}/L_{\rm rad}^{\rm iso}$ for Holmberg X II X-1 is three times larger based on the jet power reported by \citet{csehUnveilingRecurrentJets2014}.
In this case, a rotating black hole is preferred as the central object.

Regarding V404 Cygni, the ratio of kinetic to isotropic X-ray luminosities is $10^{-4}-10^{-3}$, which is much smaller than our results.
The X-ray luminosity at the outburst is about $10^{39} {\rm erg\ s^{-1}}$ \citep{mottaSwiftObservationsV4042017} and jet power is $10^{35} - 10^{36} {\rm erg\ s^{-1}}$ \citep{tetarenkoTrackingVariableJets2019} in V404 Cygni. 
This X-ray luminosity is close to the Eddington luminosity of the black hole with $10M_\odot$ \citep{shahbazMassBlackHole1994,kharghariaNearinfraredSpectroscopyLowmass2010} so that the radiation force is not strong enough to efficiently accelerate the gas.
In the case of ULX Holmberg II X-1 and ULX IC342 X-1,
the X-ray luminosity is about $10$ times higher than the Eddington luminosity, so the strong radiation can help the mass ejection.

The $L_{\rm kin}/L_{\rm rad}^{\rm iso}$ of GRS 1915+105 at high state is, on the other hand, about $20$ \citep{doneGRS19151052004,fenderGRS19151052004}, which is much higher than our results'. 
This fact indicates that the GRS 1915+105 might originate from the rapidly rotating black hole \citep{mcclintockSpinNearExtremeKerr2006}, or from an efficient energy conversion mechanism.
If the radiation energy and/or the magnetic field energy
are converted to the kinetic energy at a large distance,
the large value of $L_{\rm kin}/L_{\rm rad}^{\rm iso}$ might be explained since we measure $L_{\rm kin}$ and $L_{\rm rad}^{\rm iso}$ at $r=200r_{\rm g}$ in the present study.
The energy conversion from magnetic to kinetic energy could occur at a large distance \citep{vlahakisRelativisticMagnetohydrodynamicsApplication2003,sapountzisMRIImprintShortGRB2019a}. Also, it has been reported that the radiation energy is converted to the kinetic energy in the distant region \citep{sadowskiPowerfulRadiativeJets2015a}.
If this is the case, even the supercritical disk
is identified as an object exhibiting strong jets/winds.
If the energy conversion efficiencies from the radiation energy to the kinetic energy, and from the magnetic energy to the kinetic energy are understood, it may be possible to restrict the spin parameter from the ratio of $L_{\rm kin}/L_{\rm rad}^{\rm iso}$.
For example, 
if the conversion from radiation energy to kinetic energy is inefficient,
sources with large $L_{\rm kin}/L_{\rm rad}^{\rm iso}$ could not be explained by a radiatively driven model,
but a magnetically accelerated model would be preferred. Such a source would originate
from the rapidly rotating black hole.
For a more detailed study, the simulations with a larger computational domain are required.

The large $L_{\rm kin}/L_{\rm rad}^{\rm iso}$ may be due to observer's viewing angle
(angle between the rotation axis and the line of sight).
The X-ray luminosity estimated from the observation is approximately $L_{\rm rad}^{\rm iso}$ for a face-on observer, but is expected to decrease as the observer's viewing angle increases.
Indeed, a relatively large angle, $\sim 60-70^\circ$, is suggested by observations of GRS 1915+105
\citep{mirabelSuperluminalSourceGalaxy1994,fenderMERLINObservationsRelativistic1999,blumMEASURINGSPINGRS2009a}.
To elucidate this, it is necessary to know the large-scale structure.
The simulations with a larger computational domain are left as important future work.

\subsection{Limitations and Future Work}

We showed that the energy outflow efficiency $L_{\rm tot}/\dot{M}_{\rm in}$ of supercritical accretion disk increases with $|a^*|$. A similar result is obtained by \citet{sadowskiNumericalSimulationsSupercritical2014b}. The $L_{\rm tot}/\dot{M}_{\rm in}$ is $0.003$ for $a^*=0$ in our model is comparable to that of \citet{sadowskiNumericalSimulationsSupercritical2014b}.
For $a^*=0.9$, our result is $\sim 10$ times larger than that of \citet{sadowskiNumericalSimulationsSupercritical2014b}. This big difference would be due to an occurrence of transition from SANE to MAD in our simulation. We computed the normalized magnetic flux threading the horizon $\phi_{\rm BH}$, and confirmed that the flux exceeds the critical value $\phi_{\rm BH}=50$ at $t\sim 4500r_{\rm g}$, only for $a^* = 0.9$, before the end of the simulations. 
Thus, the disk is in MAD state and the strong jet emerges at $t \gtrsim 4500r_{\rm g}$
\citep{tchekhovskoyGeneralRelativisticModeling2012c,narayanJetsMagneticallyArrested2022}, leading to an increase of $L_{\rm tot}/\dot{M}_{\rm in}$. When the time domain of $t>4500r_{\rm g}$ is excluded, $L_{\rm tot}/\dot{M}_{\rm in}$ is $0.09$ for $a^*=0.9$. This value is consistent with that by \citet{sadowskiNumericalSimulationsSupercritical2014b}, $L_{\rm tot}/\dot{M}_{\rm in} \sim 0.02$.
We also note that the transition occurs at about $t=4500t_{\rm g}$ in our model, which is earlier than \citet{sadowskiNumericalSimulationsSupercritical2014b} $(t =12000t_{\rm g})$. We think the reason might be the plasma beta at the initial state, which is smaller in our model than in theirs. The stronger magnetic field at the initial state leads to a faster transition to MAD. We aim to study the MAD state in the supercritical accretion in the forthcoming paper, but our discussion is concentrated on mainly the SANE state in this manuscript.

Enhancement of the energy outflow efficiency via the black holes spin is also shown by GR-MHD calculations, which is used in the study of disks with very small accretion rates \citep{tchekhovskoyGeneralRelativisticModeling2012c,narayanJetsMagneticallyArrested2022}.
However, the absolute values of the energy outflow efficiencies are very different, while our results are less than $30\%$ at maximum, and their results can exceed $100\%$. 
This may be due to the difference in the mass accretion rate and/or 
magnetic field strength.
They investigate the MAD, which has strong magnetic fields, 
while our simulations investigate SANE.
The problem can be solved by performing simulations of the MAD 
with the supercritical accretion rate.
Such simulations are left as an important future work.

Regarding the radiation luminosity, $L_{\rm rad}/\dot{M}_{\rm in}$ is almost consistent, but is slightly larger (by a factor of $3$) than that of \citet{sadowskiNumericalSimulationsSupercritical2014b}. The reason why their results are slightly smaller, may be due to a higher mass accretion rate (by a factor of $10$) than ours. Most of the photons generated inside the disk are swallowed by the black hole due to the photon trapping effect. This mechanism is prominent for a higher mass accretion rate. Even though the mass accretion rate increases, the radiation luminosity does not rise so much in supercritical accretion disks. The resulting $L_{\rm rad}/\dot{M}_{\rm in}$ would be smaller for a higher mass accretion rate. Further investigation about the dependency of the efficiency on the accretion rate is needed for more details.

We start the simulations from the rotating equilibrium torus with the outer edge $r<100r_{\rm g}$ and study the accretion disk near the black hole and the gas ejection from it.
However, \citet{kitakiSystematicTwodimensionalRadiationhydrodynamic2018} showed 
by the two-dimensional RHD calculations that 
the gas is ejected from just inside the trapping radius,
which is the boundary between the supercritical and standard disks, 
$\sim (\dot{M}_{\rm in}/L_{\rm Edd}) r_{\rm g}$.
Thus, it is necessary to perform long term GR-RMHD simulations with a larger computational domain
in order to reveal overall structure of the supercritical flows.

The investigation of dependence of the initial magnetic field 
is also a future work.
Although the single loop of poloidal magnetic field is used as the initial magnetic field in the present study,
it has been reported that the magnetic flux at the event horizon differs depending on the initial magnetic field \citep{narayanGRMHDSimulationsMagnetized2012,sadowskiEnergyMomentumMass2013c,sadowskiGlobalSimulationsAxisymmetric2015}.
Moreover, we need to perform 
three-dimensional calculations 
in order to understand more realistic inflow-outflow structure,
although the two-dimensional simulations are performed in this study.
For the cases of the supercritical disks,
it has been reported that higher variability in the radiative flux appears in the two-dimensional simulations \citep{sadowskiThreedimensionalSimulationsSupercritical2016a}. Also the disk dynamo does not work in two-dimensions, so that it affects on the accretion rate and radiative efficiency. Global three-dimensional simulations can solve these problem.

We use M1 closure to approximately calculate the radiation field in this study.
It has been pointed out that the M1 closure does not give correct results 
where the radiation fields is highly anisotropic region like the vicinity of 
the rotation axis \citep{asahinaNumericalSchemeGeneral2020a}.
To solve this problem, we need to solve the radiation transfer equation 
without using the M1 closure approximation \citep{jiangALGORITHMRADIATIONMAGNETOHYDRODYNAMICS2014,jiangGlobalRadiationMagnetohydrodynamic2019,jiangSuperEddingtonAccretionDisks2019,ohsugaNUMERICALSCHEMESPECIAL2016,asahinaNumericalSchemeGeneral2020a}.

In this work, we evaluated the electron temperature $T_{\rm e}$ appeared in the opacity by assuming $T_{\rm e} = T_{\rm g}$. It is well known that this assumption is violated for low-density flows, e.g, low luminosity accretion disks (RIAF) \citep{narayanExplainingSpectrumSagittarius1995,manmotoSpectrumOpticallyThin1997,moscibrodzkaGeneralRelativisticMagnetohydrodynamical2016,moscibrodzkaUnravelingCircularPolarimetric2021}. 
The supercritical accretion disk targeted in this study is much denser than RIAF, so that the Coulomb interaction would work efficiently. The electron temperature is close to the ion temperature. In fact, we confirmed from our numerical results that the time scale of the Coulomb interaction is shorter than the dynamical time scale of the fluid in the entire computational domain, excluding the very vicinity of the rotation axis. 
The electron temperature is then determined by the balance between radiative cooling, viscous heating, adiabatic heating/cooling, and advective heating/cooling. Although ions and electrons are considered to decouple near the axis, the internal energy of the gas is sufficiently smaller than the magnetic and radiation energy. Thus, the effect of this decoupling on the dynamics would be negligible, despite the determination of the electron temperature being important to evaluate the emergent spectra. Two-temperature GR-RMHD simulation has already modelled the low luminosity disk \citep{sadowskiRadiativeTwotemperatureSimulations2017a,ryanTwotemperatureGRRMHDSimulations2018}. We leave the more detailed analysis as an important future work.

Radiation transfer calculations using the results of two-temperature simulations are also an important future work to restrict the black hole spin. 
This is because our simulations show that the strength of the Poynting flux around the rotation axis depends on the black hole spin. 
Since synchrotrons and synchrotron self-comptons are very sensitive to magnetic fields, differences in black hole spin are expected to induce differences in the spectral energy distribution. 
Polarization degree and polarization distributions, which are strongly affected by Faraday rotation and conversion, may also depend on black hole spin \citep{tsunetoePolarizationImaging872020,tsunetoePolarizationImagesAccretion2021,moscibrodzkaFaradayRotationGRMHD2017,eventhorizontelescopecollaborationFirstM87Event2021}.
Polarized radiation transfer calculations are also important future work. 

\section{Summary}\label{sec:sum}
We have performed two-dimensional axisymmetric GR-RMHD simulations 
of the supercritical flows in the SANE state around stellar mass black holes with $a^*$ varying from $-0.9$ to $0.9$. 
Our simulations show optically and geometrically thick disks near the equatorial plane and the powerful gas ejection from the disk surface in all models.
The gas ejection is mainly induced by the radiative force, 
but magnetic force also works to accelerate when $|a^*|$ is large.

The energy outflow efficiency,
$\langle L_{\rm tot} \rangle/\langle \dot{M}_{\rm in}\rangle $,
is larger for rotating black holes than for the non-rotating black hole. 
In particular, this is found to be larger for the prograde models. 
We find $\langle L_{\rm tot} \rangle/\langle \dot{M}_{\rm in}\rangle 
=0.7\%$ for $a^*=-0.7$, $0.3\%$ for $a^*=0$, and $5\%$ for $a^*=0.7$ for $\dot{M}_{\rm in} \sim 100L_{\rm Edd}$.
Although the disks around the non-rotating black holes release the energy mainly by radiation, 
the Poynting power is enhanced more than 
the other luminosities in the case of a rotating black hole.

The power ratio of the kinetic and isotropic radiative luminosity $L_{\rm kin}/L_{\rm rad}^{\rm iso}$ tends to be larger the case for rotating the black hole than the case for not rotating black hole.
This result suggests that objects with large $L_{\rm kin}/L_{\rm rad}^{\rm iso}$ ($\gtrsim 0.15$), such as ULX IC342 X-1, may have a rapidly rotating black hole.
On the other hand, a slowly rotating or non-rotating black hole would be suitable for an central object for ULXs with small $L_{\rm kin}/L_{\rm rad}^{\rm iso}$  ($\lesssim 0.15$), such as ULX Holmberg II X-1.
For more detailed comparison with observations, GR-RMHD simulations with larger computational domain and radiation transfer simulations are needed.

\begin{acknowledgments}
Numerical computations were performed with computational resources provided by the Multidisciplinary Cooperative Research Program in the Center for Computational Sciences, University of Tsukuba, Oakforest-PACS operated by the Joint Center for Advanced High-Performance Computing (JCAHPC) and, Cray XC50 at
the Center for Computational Astrophysics (CfCA) of the National Astronomical Observatory of Japan (NAOJ).
This work was supported by JST, the establishment of university fellowships towards the creation of science technology innovation, Grant Number JPMJFS2106(A.U.).
This work was also supported by JSPS KAKENHI Grant Numbers 
21H04488, 18K03710(K.O.), 20K11851, 20H01941, 20H00156 (H.R.T), 18K13591(Y.A.)
A part of this research has been funded by the MEXT as ”Program for Promoting Researches on the Supercomputer Fugaku”
(Toward an unified view of the universe: from large scale structures to planets, JPMXP1020200109) (K.O., H.R.T., Y.A., and A.U.), and by Joint Institute for Computational Fundamental Science (JICFuS, K.O.). 

\end{acknowledgments}


\bibliographystyle{aasjournal}
\bibliography{research2019-2021}



\end{CJK}
\end{document}